\documentclass[conference]{IEEEtran}
\IEEEoverridecommandlockouts

\usepackage{cite}
\usepackage{amsmath,amssymb,amsfonts}
\usepackage{algorithmic}
\usepackage{graphicx}
\usepackage{textcomp}
\usepackage{xcolor}

\def\BibTeX{{\rm B\kern-.05em{\sc i\kern-.025em b}\kern-.08em
    T\kern-.1667em\lower.7ex\hbox{E}\kern-.125emX}}
\begin{document}

\title{Time-Complexity Characterization of NIST Lightweight Cryptography Finalists
}

\author{\IEEEauthorblockN{Najmul Hasan}
\IEEEauthorblockA{\textit{Department of Mathematics \& Computer Science} \\
\textit{University of North Carolina}\\
Pembroke}
\and
\IEEEauthorblockN{Prashanth BusiReddyGari}
\IEEEauthorblockA{\textit{Department of Mathematics \& Computer Science} \\
\textit{University of North Carolina}\\
Pembroke}

}


\maketitle

\IEEEpubidadjcol


\begin{abstract}
Lightweight cryptography is becoming essential as emerging technologies in digital identity systems and Internet of Things verification continue to demand strong cryptographic assurance on devices with limited processing power, memory, and energy resources. As these technologies move into routine use, they demand cryptographic primitives that maintain strong security and deliver predictable performance by clear theoretical models of time complexity. Although NIST’s lightweight cryptography project provides empirical evaluations of the ten finalist algorithms, a unified theoretical understanding of their time-complexity behavior remains absent. This work introduces a symbolic model that decomposes each scheme into initialization, data-processing, and finalization phases, enabling formal time-complexity derivation for all ten finalists. 
The results clarify how design parameters shape computational scaling on constrained mobile and embedded environments. The framework provides a foundation needed to distinguish algorithmic efficiency and guides the choice of primitives capable of supporting security systems in constrained environments.

\end{abstract}



\section{Introduction}
Internet of Things (IoT) devices and embedded systems operate under severe resource constraints, requiring cryptographic solutions that balance security with minimal computational overhead \cite{williams2022survey}. Traditional cryptographic algorithms prioritize security strength but often demand processing power and memory beyond the capabilities of resource-limited devices. Lightweight cryptography (LWC) addresses this gap by optimizing algorithms for constrained environments where energy consumption, processing speed, and memory usage directly impact operational viability.

The National Institute of Standards and Technology (NIST) initiated a standardization process to evaluate and select cryptographic algorithms specifically designed for resource-constrained applications \cite{turan2023status}. Following multiple evaluation rounds, ten finalist schemes emerged from this competition, each employing distinct design strategies and operational characteristics: Ascon \cite{ascon2021nist}, Elephant \cite{beyne2021elephant}, GIFT-COFB \cite{banik2021}, Grain-128AEAD \cite{hell2021grain}, ISAP \cite{dobraunig2021isap}, PHOTON-Beetle \cite{photon-beetle}, Romulus \cite{Romulusv1.3}, SPARKLE (Schwaemm and Esch) \cite{Beierle2021}, TinyJAMBU \cite{wu2021tinyjambu}, and Xoodyak \cite{Daemen2021Xoodyak}. In February 2023, NIST selected Ascon as the winner for standardization \cite{konstantopoulou2025review}, though comprehensive analysis of all finalists remains valuable for understanding design trade-offs \cite{madushan2022review}.

Time complexity directly determines algorithm feasibility in real-world deployments \cite{radhakrishnan2024efficiency}. For lightweight cryptography, where devices operate on minimal power budgets, computational efficiency translates to extended battery life and improved system responsiveness. Quantifying time complexity across encryption, decryption, and associated data processing phases enables meaningful performance comparisons among competing schemes. Recent hardware and software implementations have demonstrated significant performance variations across platforms \cite{sorescu2025comparative}, reinforcing the need for systematic complexity analysis \cite{sarker2025systematic}.

This analysis derives symbolic time complexity expressions for all ten NIST finalists using mathematical models grounded in their algorithmic specifications. Section 2 establishes the theoretical foundation for complexity analysis in cryptographic contexts. Section 3 presents the mathematical framework. Section 4 provides detailed complexity analyses and comparative evaluation of each algorithm. Section 5 concludes with research directions.


\section{Background on Complexity Analysis}
Complexity analysis provides a structured framework for quantifying algorithmic computational requirements. Originating from theoretical computer science, this methodology has become essential for cryptographic systems, where balancing security with efficiency determines practical deployability. In resource-constrained environments typical of lightweight cryptography, time complexity directly influences algorithm feasibility, as computational overhead translates to energy consumption and processing latency.

Time complexity quantifies the computational effort required to process input of given size. For cryptographic schemes, this metric identifies bottlenecks in encryption, decryption, and data processing phases, ensuring compatibility with devices operating under minimal power and processing constraints \cite{williams2022survey, radhakrishnan2024efficiency}. The measurement of time complexity enables systematic performance comparison across algorithms, revealing efficiency trade-offs inherent in different cryptographic designs \cite{el2023analysis}.

Despite widespread recognition of time complexity's importance, systematic symbolic analyses remain limited in lightweight cryptography research. NIST's standardization process emphasized performance through empirical benchmarks measuring execution time and throughput across platforms \cite{turan2023status, konstantopoulou2025review}. However, official documentation for finalists such as Ascon \cite{ascon2021nist}, ISAP \cite{dobraunig2021isap}, and TinyJAMBU \cite{wu2021tinyjambu} focuses on implementation optimizations without deriving formal complexity expressions. This absence of theoretical frameworks creates significant opportunity for unified complexity evaluation methodologies \cite{thakor2021lightweight}.

This research addresses the gap by developing comprehensive time complexity analysis using symbolic mathematical models. The framework derives expressions for core cryptographic operations, enabling thorough examination of computational trade-offs inherent in each algorithm and providing insights for deployment in resource-constrained environments.


\section{Time Complexity Model}

Evaluating time complexity in lightweight cryptography requires systematic analysis of computational requirements across operational phases. The total time complexity is modeled as the sum of initialization, data processing, and finalization complexities:

\begin{figure}[h]
    \centering
    \includegraphics[width=\columnwidth]{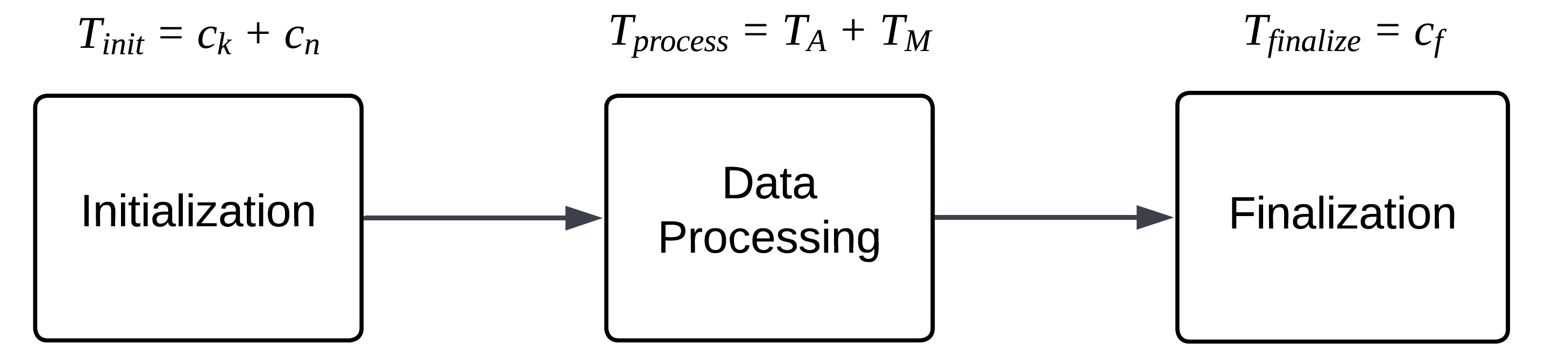}
    \caption{Time complexity model showing the Initialization ($T_{\text{init}}$), Data Processing ($T_{\text{process}}$), and Finalization ($T_{\text{finalize}}$) phases with their equations.}
    \label{fig:time_complexity}
\end{figure}

\begin{equation}
T_{\text{total}} = T_{\text{init}} + T_{\text{process}} + T_{\text{finalize}}
\end{equation}

where $T_{\text{init}}$, $T_{\text{process}}$, and $T_{\text{finalize}}$ represent initialization, data processing, and finalization phase complexities, respectively. This three-phase decomposition (Fig.~\ref{fig:time_complexity}) enables systematic analysis across diverse algorithmic structures.

The initialization phase establishes internal state through key and nonce setup operations independent of input size:

\begin{equation}
T_{\text{init}} = c_k + c_n
\end{equation}

where $c_k$ and $c_n$ denote fixed computational costs for key and nonce initialization. Algorithms such as TinyJambu \cite{wu2021tinyjambu} perform XOR operations during this phase with complexity $O(3 \cdot b_{640})$.

Data processing handles associated data and message blocks through padding, permutations, and encryption operations. Total processing complexity decomposes as:

\begin{equation}
T_{\text{process}} = T_A + T_M
\end{equation}

where $T_A = a \cdot (T_p + c_A)$ and $T_M = m \cdot (T_p + c_M)$ represent associated data and message processing complexities. Here, $a$ and $m$ denote block counts, $T_p$ represents permutation time, and $c_A$, $c_M$ account for operation overhead. PHOTON-Beetle \cite{photon-beetle} exhibits complexity $O(\lceil \ell_A / r \rceil \cdot b)$, while Xoodyak \cite{Daemen2021Xoodyak} contributes $O(\lceil |M|/r \rceil \cdot b)$.

Finalization ensures data integrity through authentication tag generation or validation with input-independent complexity:

\begin{equation}
T_{\text{finalize}} = c_f
\end{equation}

where $c_f$ denotes constant tag computation time. GIFT-COFB \cite{banik2021} achieves $O(1)$ finalization complexity.

This framework accommodates structural diversity across lightweight cryptographic algorithms. Permutation-based designs like Ascon \cite{ascon2021nist} and Xoodyak \cite{Daemen2021Xoodyak} exhibit complexity dominated by $T_p$, while stream ciphers such as Grain-128AEAD \cite{hell2021grain} introduce keystream generation impacting both $T_A$ and $T_M$. Block ciphers including TinyJambu \cite{wu2021tinyjambu} and Romulus \cite{Romulusv1.3} show higher $T_p$ from iterative encryption rounds. SPARKLE \cite{Beierle2021} demonstrates overall complexity:

\begin{equation}
\small
T_{\text{overall}} = O\left( 2{\cdot}\left\lceil \frac{|A|}{r} \right\rceil{\cdot}b + 3{\cdot}\left\lceil \frac{|M|}{r} \right\rceil{\cdot}b + \left\lceil \frac{d}{r} \right\rceil{\cdot}b + 2{\cdot}b \right)
\end{equation}

This abstraction enables platform-independent algorithm evaluation, facilitating comparative analysis and identifying bottlenecks for resource-constrained deployments.


\section{Algorithm Complexity Analysis}

This section presents the time complexity analysis of the ten NIST finalist lightweight cryptographic algorithms by applying the three-phase model from Section III. We derive symbolic complexity expressions for each algorithm's encryption and decryption operations, quantifying computational requirements as functions of message length, associated data length, and algorithm-specific parameters. Table~\ref{tab:lwc-comparison} summarizes the derived complexities for all finalists.

\begin{table*}[t]
\renewcommand{\arraystretch}{1.6}
\caption{Time complexities of NIST LWC finalists. \\
Parameters: $b$ = permutation cost, $r$ = rate, \\ $n$ = block size, $P$ = permutation complexity, $d$ = digest length.}
\label{tab:lwc-comparison}
\vspace{0.2cm}
\setlength{\tabcolsep}{8pt}
\centering
\small
\begin{tabular}{|l|c|}
\hline
\textbf{Algorithm} & \textbf{Time Complexity} \\ \hline
ASCON            & $O(l_A \cdot b + l_P \cdot b)$      \\ \hline
Elephant         & $O(\ell_M \cdot P + \ell_A \cdot P)$      \\ \hline
GIFT-COFB        & $O(\ell_A + \ell_M)$                     \\ \hline
Grain-128AEAD    & $O(|M| + |AD|)$                    \\ \hline
ISAP             & $O(|A| + |M|)$                     \\ \hline
PHOTON-Beetle    & $O\left(\left\lceil\frac{|A|}{r}\right\rceil \cdot b + \left\lceil\frac{|M|}{r}\right\rceil \cdot b\right)$ \\ \hline
Romulus-N        & $O\left(\frac{|A|}{n} \cdot b + \frac{|M|}{n} \cdot b\right)$ \\ \hline
SPARKLE          & $O\left(2 \cdot \frac{|A|}{r} \cdot b + 3 \cdot \frac{|M|}{r} \cdot b + \frac{d}{r} \cdot b + 2b\right)$ \\ \hline
TinyJambu        & $O\left(6b_{640} + 2 \cdot \frac{|A|}{32} \cdot b_{640} + 2 \cdot \frac{|M|}{32} \cdot b_{1024} + 4b_{1024}\right)$ \\ \hline
Xoodyak          & $O\left(2b + 2 \cdot \frac{|A|}{r_{\text{in}}} \cdot b + 2 \cdot \frac{|P|}{r_{\text{out}}} \cdot b\right)$ \\ \hline
\end{tabular}
\end{table*}

The complexities exhibit linear scaling with input lengths, with coefficients determined by algorithm-specific parameters including state size, permutation design, and mode of operation. The expressions reveal fundamental differences in computational scaling that reflect distinct design philosophies optimized for specific deployment scenarios.

\subsection{Permutation-Based Algorithms}

Five algorithms employ sponge-based or permutation-centered constructions. ASCON \cite{ascon2021nist} achieves complexity $O(l_A \cdot b + l_P \cdot b)$ through 12-round initialization followed by iterative processing of associated data and plaintext blocks. Its streamlined design minimizes overhead, achieving among the most efficient scaling profiles in the finalist set and making it particularly suitable for resource-constrained deployments.

PHOTON-Beetle \cite{photon-beetle} demonstrates similar scaling $O(\lceil|A|/r\rceil \cdot b + \lceil|M|/r\rceil \cdot b)$ using the PHOTON256 permutation. The ceiling functions reflect block padding overhead, while the rate parameter $r$ determines processing granularity. Xoodyak \cite{Daemen2021Xoodyak} employs the Cyclist mode over the Xoodoo permutation, incorporating distinct rates for absorbing ($r_{\text{in}}$) and squeezing ($r_{\text{out}}$) phases. This flexibility enables optimization for specific throughput requirements while maintaining complexity comparable to ASCON and PHOTON-Beetle.

Elephant \cite{beyne2021elephant} leverages LFSR-based masking with three variants (Dumbo, Jumbo, Delirium) using different permutations (Spongent-$\pi$[160], Spongent-$\pi$[176], Keccak-$f$[200]). Its complexity $O(\ell_M \cdot P + \ell_A \cdot P)$ scales with permutation cost $P$, which varies by variant. SPARKLE \cite{Beierle2021} employs ARX-based permutations for both SCHWAEMM (AEAD) and ESCH (hashing), with complexity $O(2 \cdot |A|/r \cdot b + 3 \cdot |M|/r \cdot b + d/r \cdot b + 2b)$. The additional terms reflect enhanced functionality supporting both encryption and hashing operations, distinguishing it from single-purpose designs.

\subsection{Block Cipher-Based Algorithms}

Three algorithms employ block cipher or register-based constructions with varying complexity characteristics. GIFT-COFB \cite{banik2021} achieves the notably simple complexity $O(\ell_A + \ell_M)$ through the COFB feedback mode with the lightweight GIFT cipher. This linear scaling without multiplicative factors represents the simplest expression among all finalists, making GIFT-COFB exceptionally efficient for large messages where the absence of per-block multipliers reduces computational overhead.

Romulus-N \cite{Romulusv1.3} uses tweakable block ciphers with complexity $O(|A|/n \cdot b + |M|/n \cdot b)$, where $n$ denotes the block size. The tweakable construction provides additional security properties at the cost of block cipher invocations proportional to message and associated data lengths. TinyJambu \cite{wu2021tinyjambu} employs a 128-bit NFSR with dual permutations ($P_{640}$ and $P_{1024}$) for different phases. Its complexity $O(6b_{640} + 2 \cdot |A|/32 \cdot b_{640} + 2 \cdot |M|/32 \cdot b_{1024} + 4b_{1024})$ includes fixed initialization costs and separate processing costs for associated data (using $P_{640}$) and messages (using $P_{1024}$). The use of 32-bit blocks enables finer processing granularity compared to larger-block designs, though this incurs higher per-block overhead that becomes proportionally less significant for longer messages.

\subsection{Stream Cipher Algorithm}

Grain-128AEAD \cite{hell2021grain} stands as the sole stream cipher finalist, combining NFSR and LFSR components for keystream generation. Its complexity $O(|M| + |AD|)$ scales directly with bit count rather than block count, eliminating padding overhead entirely. The multi-phase initialization process ensures thorough state mixing through state setup, key reintroduction, and register initialization. This bit-level processing architecture offers particular advantages in scenarios requiring fine-grained operations without block alignment constraints, complementing the block-oriented approaches of other finalists.

\subsection{Hybrid Algorithm}

ISAP \cite{dobraunig2021isap} combines sponge-based encryption with session key derivation, achieving complexity $O(|A| + |M|)$. The key derivation step, performed with constant overhead, generates ephemeral encryption keys for each session. This hybrid approach provides resistance against side-channel attacks through key separation while maintaining efficient linear scaling with input lengths comparable to GIFT-COFB and Grain-128AEAD.

The ten finalists demonstrate diverse approaches to lightweight cryptography, with complexity expressions reflecting fundamental trade-offs between simplicity, functionality, and security properties. Algorithms exhibit different scalability characteristics: GIFT-COFB, Grain-128AEAD, and ISAP achieve the simplest linear scaling with minimal overhead, while SPARKLE and TinyJambu incorporate additional constant terms reflecting enhanced functionality or security margins. Block-based processing introduces granularity effects, where algorithms using smaller blocks process data in finer increments but may incur higher per-block overhead, while larger-block designs reduce iteration counts at the cost of increased padding for small messages. Architectural choices drive complexity coefficients, with permutation-based designs concentrating computational cost in the permutation function itself (scaling as $O(\ell \cdot b)$ where $b$ represents permutation cost), while block cipher-based designs similarly scale with cipher invocation cost. These characteristics demonstrate that no single algorithm dominates across all metrics, enabling selection based on specific deployment constraints including minimal overhead (GIFT-COFB), bit-level processing (Grain-128AEAD), side-channel resistance (ISAP), or integrated hashing (SPARKLE).


\section{Conclusion}

This work introduced a unified symbolic framework for analyzing the time complexity of the ten NIST LWC finalists and clarifies how their design choices influence computational cost on constrained platforms. The analysis provides a coherent theoretical basis for comparing these algorithms and identifying candidates that align with the demands of emerging identity technologies. Building on these results, our work focuses on validating the theoretical complexity predictions within mobile drivers license environments and related digital identity operations. Extending the framework into practical evaluation will offer deeper insight into lightweight cryptography performance in real systems that rely on predictable computation under resource constraints.


\bibliographystyle{IEEEtran}
\bibliography{references}


\end{document}